\documentclass{article}

\usepackage{arxiv}
\usepackage[utf8]{inputenc}
\usepackage[T1]{fontenc}
\usepackage{amsmath,amssymb}
\usepackage{graphicx}
\usepackage{booktabs,tabularx,array}
\usepackage{microtype}
\usepackage{courier}
\usepackage{cite}
\usepackage{xurl}
\usepackage[hidelinks]{hyperref}
\usepackage{listings}

\newcolumntype{Y}{>{\raggedright\arraybackslash}X}
\lstdefinestyle{prompt}{
  basicstyle=\small\ttfamily,
  breaklines=true,
  columns=fullflexible,
  keepspaces=true,
  showstringspaces=false,
  frame=single,
  framerule=0.3pt,
  xleftmargin=0.5em,
  xrightmargin=0.5em,
  aboveskip=0.7em,
  belowskip=0.7em
}

\title{NoveltyScope: A Multi-Agent Research Assistant for Evidence-Grounded Novelty Assessment of Scientific Papers}

\author{
  Yueran Tian\\
  School of Computing \\
  Newcastle University\\
  Newcastle upon Tyne, UK\\
  \texttt{Everest\_Tian@outlook.com}
  \And
  Huizhi Liang\\
  School of Computing \\
  Newcastle University\\
  Newcastle upon Tyne, UK\\
  \texttt{Huizhi.Liang@newcastle.ac.uk}
}

\date{}
\renewcommand{\shorttitle}{NoveltyScope: Evidence-Grounded Novelty Assessment}
\hypersetup{
  pdftitle={NoveltyScope: A Multi-Agent Research Assistant for Evidence-Grounded Novelty Assessment of Scientific Papers},
  pdfauthor={Yueran Tian, Huizhi Liang},
  pdfsubject={Evidence-grounded scientific novelty assessment},
  pdfkeywords={scientific novelty, scholarly retrieval, multi-agent evaluation, research assistance}
}

\begin{document}
\raggedbottom
\maketitle
\begin{abstract}
  NoveltyScope is a prototype that compares a scientific paper with retrieved prior work and produces a structured report and a five-level novelty score.
The system extracts the core task and at most two author-claimed contributions, then searches a closed local corpus using BM25 and dense retrieval with reciprocal rank fusion.
Contribution comparisons request paired passages for potential refutations, and quotations are checked against cached source text.
Four scoring roles (Prosecutor, Defender, Judge, and Critic) use the resulting evidence, with at most one revision of the Judge's score.
Evaluation uses 50 class-balanced RINoBench samples whose idea-level labels are retained for full-paper input.
Mean absolute error is 1.48 and 1.26 for single-LLM and multi-agent scoring with Gemini, respectively, and 1.22 and 1.16 with Qwen3.
The vector-similarity baseline has an error of 1.68.
A HiCLIP case study illustrates the contribution comparisons and paired evidence passages.
Overall, the results demonstrate the feasibility of evidence-grounded novelty assessment and show that the multi-agent configurations achieve lower scoring errors than their single-LLM counterparts across both evaluated models.

\end{abstract}

\section{Introduction}
\label{sec:introduction}
Assessing a paper's novelty requires comparing its claimed contributions with relevant prior work.
Peer-review workloads are substantial and unevenly distributed.
A modelling study of biomedical peer review estimated that 20\% of researchers performed 69--94\% of reviews~\cite{repec:plo:pone00:0166387}; a separate estimate placed global reviewing time in 2020 above 100 million hours~\cite{Aczel2021PeerReviewCost}.
Literature-aware novelty assessment makes these comparisons explicit and links judgements to evidence~\cite{afzal-etal-2026-beyond}.
Judgement also varies between reviewers: a study of the same NIH grant applications found low inter-reviewer agreement~\cite{article}.

A retrieval ranking identifies related papers, but leaves the technical comparison to the reader.
Papers addressing the same task may use different methods, and papers using different terminology may describe a similar mechanism.
A novelty assessment must identify the relevant overlap or difference and explain how it bears on the claimed contribution.
The source passages matter because a plausible model-generated explanation may misrepresent the work it cites.

NoveltyScope takes a paper PDF or URL as input and returns a structured report and a score from 1 to 5.
It extracts the core task and at most two author-claimed contributions, then retrieves related papers from a closed local collection through separate task and contribution queries.
The report contains the retrieved references, technical comparisons, and paired passages for potential refutations.
A four-role scoring stage uses these records to argue for lower or higher novelty, assign a score, and check its use of the rubric.
The Critic can request one revision of the Judge's decision.

Restricting retrieval to a local collection fixes the documents that can enter the comparison.
Recording their identifiers and checking quoted passages allows a reader to examine the sources behind a judgement.
The assessment is conditional on that collection's coverage and on the claims selected from the target paper.

\begin{samepage}
The paper describes two system-level contributions:
\begin{enumerate}
  \item A reporting pipeline that retrieves prior work at task and contribution levels, compares it with the target paper, and checks quoted passages against cached source text.
  \item A four-role scoring workflow that records arguments for lower and higher novelty, explains the selected score relative to adjacent labels, and permits one rubric-based revision.
\end{enumerate}
\end{samepage}
Both use existing retrieval and language-model components to organise the comparison and scoring process.

Evaluation comprises a 50-sample, class-balanced RINoBench comparison and an illustrative HiCLIP report.
Mean absolute error (MAE) is lower for multi-agent scoring in both tested backends: 1.26 versus 1.48 for Gemini and 1.16 versus 1.22 for Qwen3.
The scoring prompts differ between conditions, and the benchmark's idea-level labels are applied to full papers.
These choices limit attribution of the differences to the agent architecture.
NoveltyScope is intended for reviewers who need to inspect and contest the evidence behind a score.

\section{Related Work}
\label{sec:related}

\subsection{Scientific Novelty Assessment}
Scientific novelty is related to, but distinct from, originality, creativity, innovation, and breakthrough~\cite{repec:spr:scient:v:130:y:2025:i:2:d:10.1007_s11192-025-05234-0}.
New combinations of familiar components are one relevant form: Uzzi et al.\ associate scientific impact with atypical combinations embedded in a conventional knowledge base~\cite{doi:10.1126/science.1240474}.
Question--method combinations~\cite{repec:eee:infome:v:16:y:2022:i:2:s1751157722000347}, integrated topic modelling~\cite{WANG2024101587}, and changes in scholarly networks~\cite{AMPLAYO2018542} provide different operationalisations of novelty.

Wu et al.\ combine human review knowledge with LLM-based representations of methodology for novelty evaluation~\cite{Wu_2025_novelty}.
Wang et al.\ report delayed recognition of novel research, which limits the use of short citation windows for early assessment~\cite{WANG20171416}.
Liu and Zhai's axiomatic evaluation finds that existing novelty metrics do not consistently satisfy all proposed axioms~\cite{liu2026axiomaticbenchmarkevaluationscientific}.

SchNovel contains 15,000 paper pairs from six arXiv fields, and its associated RAG-Novelty method evaluates papers in retrieved context~\cite{lin2024evaluatingenhancinglargelanguage}.
Its labels rely on the chronological assumption that the later paper is more novel.
RINoBench instead supplies five-level novelty labels for research ideas derived from expert review information~\cite{schopf2026ideanovelautomatedbenchmark}.
NoveltyScope uses the RINoBench scale with full-paper inputs; Section~\ref{sec:data} describes this label adaptation.

\subsection{Evidence-Grounded Research Assistance}
LLMs have been applied to literature review, scientific reasoning, and other research-assistance tasks~\cite{luo2025llm4srsurveylargelanguage}.
Agent Laboratory investigates LLM agents as research assistants~\cite{schmidgall2025agentlaboratoryusingllm}, while OpenScholar combines scientific retrieval with literature-grounded synthesis~\cite{asai2024openscholarsynthesizingscientificliterature}.
SciEval~\cite{sun2024scievalmultilevellargelanguage} and ResearchBench~\cite{liu2025researchbenchbenchmarkingllmsscientific} examine scientific capabilities, while work on practical LLM grounding identifies evaluation and diagnostic challenges~\cite{Kenthapadi_2024}.
Zhang et al.\ examine LLM use across the scientific method, including its methodological constraints~\cite{zhang2025exploring}.

Retrieval-augmented generation connects a language model to external documents for knowledge-intensive tasks~\cite{lewis2021retrievalaugmentedgenerationknowledgeintensivenlp}.
A survey of retrieval-augmented generation discusses factual grounding, knowledge limitations, and failures in both retrieval and generation~\cite{gao2024retrievalaugmentedgenerationlargelanguage}.
For scholarly search, LitSearch shows that realistic queries require understanding research concepts beyond matching keywords~\cite{ajith-etal-2024-litsearch}.
SPECTER develops citation-informed document representations to capture relationships between scientific papers~\cite{cohan2020specterdocumentlevelrepresentationlearning}.
In NoveltyScope, retrieval selects candidates for technical comparison; scoring follows the comparison stage.

OpenNovelty also extracts contribution claims, retrieves related work, compares tasks and contributions, and generates an evidence-linked report~\cite{zhang2026opennoveltyllmpoweredagenticverifiable}.
Its use of WisPaper for scholarly retrieval~\cite{ju2026wispaperaischolarsearch} contrasts with NoveltyScope's fixed local collection.
GraphMind presents paper components and their relationships to the literature as an interactive graph~\cite{dasilva2025graphmindinteractivenoveltyassessment}.
NoveltyScope combines a local retrieval corpus with a four-role procedure for assigning a five-level score to the resulting comparisons.
The quantitative comparison is restricted to scoring configurations within NoveltyScope.

\subsection{Multi-Agent Evaluation}
Multi-agent language-model workflows distribute tasks through roles and communication rules.
CAMEL uses role-based interaction~\cite{li2023camelcommunicativeagentsmind}, MetaGPT encodes standard operating procedures~\cite{hong2024metagptmetaprogrammingmultiagent}, and AutoGen organises applications around agent conversation~\cite{wu2023autogenenablingnextgenllm}.
Communication-centred analysis emphasises that interaction protocols and objectives shape the resulting reasoning~\cite{yan2025selftalkcommunicationcentricsurveyllmbased}.

Task allocation, inter-agent reasoning, context, and memory remain challenges~\cite{han2026llmmultiagentsystemschallenges}.
LLM-Coordination identifies limitations in joint planning and reasoning about collaborators~\cite{agashe2025llmcoordinationevaluatinganalyzingmultiagent}.
Communication can also add substantial token costs, motivating economical message exchange~\cite{zhang2024cutcrapeconomicalcommunication}, while AgentVerse examines beneficial and harmful behaviours emerging from collaboration~\cite{chen2023agentversefacilitatingmultiagentcollaboration}.
NoveltyScope uses a fixed role sequence, a shared evidence payload, and at most one score revision.
This design limits the number of exchanges, while its evaluation compares the scoring configurations as a whole.

\section{Method}
\label{sec:method}

\subsection{Task and Overview}
\label{sec:task}
Given a target paper $x$ and related documents $D=\{d_1,\ldots,d_n\}$ retrieved from a local corpus $\mathcal{C}$, NoveltyScope produces a discrete novelty score $y$ and a structured report $r$:
\begin{equation}
  f:(x,D)\mapsto(y,r),\qquad y\in\{1,2,3,4,5\}.
  \label{eq:task}
\end{equation}
The score is conditional on the retrieved documents and the contributions extracted from $x$.
The report records the target's core task, claimed contributions, related papers, comparisons, and supporting passages.
Its structured records also supply the input to the scoring stage.

\begin{figure}[htbp]
  \centering
  \includegraphics[width=\textwidth]{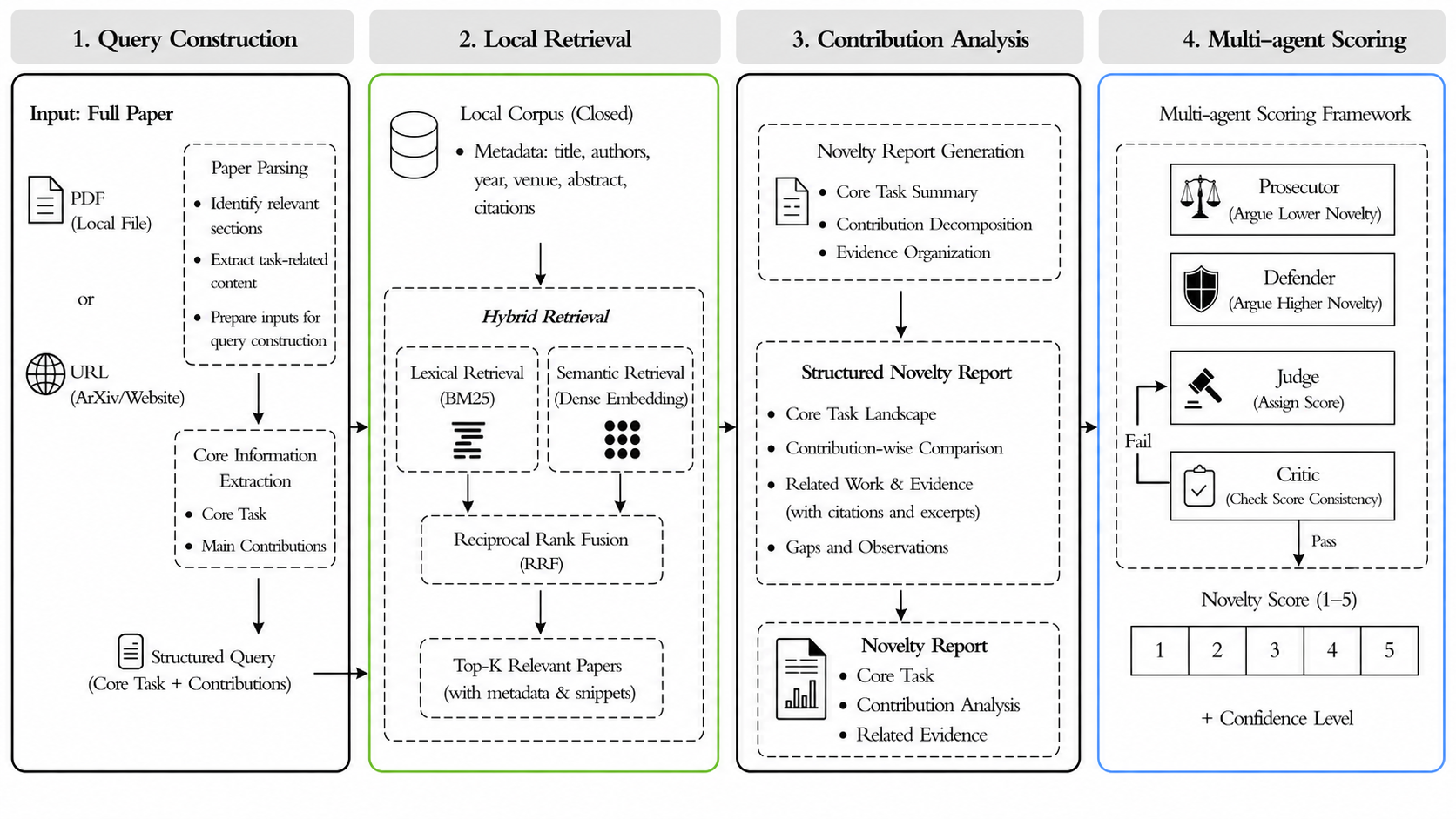}
  \caption{NoveltyScope pipeline. Task- and contribution-level queries retrieve papers from a closed local corpus; structured comparisons supply the novelty report and scoring evidence. The Critic-to-Judge return path permits at most one revision, without new Prosecutor or Defender arguments.}
  \label{fig:pipeline}
\end{figure}

Figure~\ref{fig:pipeline} shows the four stages: query construction, local retrieval, evidence comparison and report generation, and multi-agent scoring.
Task and contribution queries are formed and searched separately.
Retrieved papers are then grouped by core task and compared with the target's claims.
The scoring stage assigns an ordinal label using these comparisons.

\subsection{Paper Representation and Query Construction}
\label{sec:queries}
The input is a local PDF or a paper URL, from which the prototype obtains unstructured text.
The abstract, introduction, and conclusion form the context for task and contribution extraction.
The contribution-extraction prompt also permits the title as a source.
The core task is represented by a short statement of the central research problem, excluding performance results and implementation details.

The contribution extractor returns at most two explicitly author-claimed contributions and excludes claims consisting only of numerical performance improvements.
Each contribution contains a name of at most eight words, a verbatim author-claim span of at most 15 words, a one-sentence description of at most 20 words, and a source hint identifying the title, abstract, introduction, or conclusion.
The prompt instructs the extractor to merge duplicate statements across sections and return fewer than two contributions when necessary.
The author-claim span and source hint link each selected contribution to the paper text.
These selected claims determine which contributions enter the comparison.

For core-task retrieval, the system uses the task statement and generated query variants.
For each contribution, it constructs a prior-work query and combines it with the author-claim text, contribution description, and contribution-level variants.
The prior-work query describes the intervention in general technical terms, avoiding method names specific to the target paper.
Task queries seek work on the same research problem; contribution queries seek earlier uses of the claimed intervention.
Variants provide alternative phrasings, and duplicate queries are removed before retrieval scores are aggregated.

\subsection{Closed-Corpus Hybrid Retrieval}
\label{sec:retrieval}
The local corpus contains related-work papers prepared for the evaluation samples.
Each record stores a paper identifier, title, abstract, retrieval-source metadata, and a PDF link.
Only titles and abstracts are indexed.
PDF links are stored as metadata and used to obtain full text for the subsequent comparisons.
An overlap described only in a method section can therefore be missed at retrieval time.

The lexical channel uses BM25~\cite{10.1561/1500000019} to rank corpus records by term overlap with a query.
The dense channel uses the SentenceTransformer model \texttt{all-MiniLM-L6-v2} to encode query text and corpus titles and abstracts.
Corpus embeddings are precomputed and L2-normalised; query embeddings are normalised at retrieval time.
The two channels combine exact task and method terms with similarity between text embeddings.
Both channels return ranked document identifiers with their associated metadata.

Raw BM25 and dense similarity scores have different scales, so the system combines ranks using weighted reciprocal rank fusion (RRF).
For query $q$ and document $d$, the fused score is
\begin{equation}
  s_{\mathrm{RRF}}(d\mid q)
  =\sum_{m\in\{\mathrm{BM25},\mathrm{Dense}\}}
    \frac{w_m}{c+\operatorname{rank}_m(d,q)},
  \label{eq:rrf}
\end{equation}
where $\operatorname{rank}_m(d,q)$ is the position under retriever $m$.
The default weights are $w_{\mathrm{BM25}}=w_{\mathrm{Dense}}=1$, with fusion constant $c=60$.
For the deduplicated query set $Q$ belonging to a task or contribution, the final score is
\begin{equation}
  s(d)=\sum_{q\in Q}s_{\mathrm{RRF}}(d\mid q).
  \label{eq:queryfusion}
\end{equation}
Documents are sorted by $s(d)$ and the highest-ranked candidates are retained for analysis.
Candidate records are normalised and merged into a common reference index.
Document identifiers, ranking positions, and retrieval sources are retained so that comparisons can be traced to the selected papers.

\subsection{Evidence Comparison and Report Generation}
\label{sec:report}
Core-task candidates are organised into a hierarchical taxonomy using their titles and abstracts.
The taxonomy prompt asks the model to group papers by research approach, question, or study context, with inclusion and exclusion rules for each non-root category.
It specifies that each supplied identifier should appear once among the leaves and that the target paper should be assigned to a leaf.
The resulting tree selects papers for core-task comparison.

Core-task comparison first considers papers in the same leaf as the target.
When no direct sibling is available, the workflow falls back to neighbouring subtopics under the same parent.
The comparison describes the shared focus, overlapping aspects, and technical distinctions, and flags likely duplicate or variant papers for manual checking.
Core-task comparisons are brief; contribution analysis requests the more detailed passage pairs described below.

For each extracted contribution, a language model compares retrieved candidates with the target using the available full-text context.
It assigns one of three statuses: \texttt{can\_refute}, \texttt{cannot\_refute}, or \texttt{unclear}.
A \texttt{can\_refute} decision requests an explanation and paired verbatim passages from the target and candidate, with paragraph or section labels and a rationale connecting the passages.
The prompt limits each quoted passage to 90 words.
A \texttt{cannot\_refute} decision instead requests a short explanation of the technical difference; \texttt{unclear} records insufficient detail for a judgement.

Supporting passages are checked against locally cached source text.
When a quoted passage cannot be aligned with that text, the associated claim is retained only as a qualified remark and excluded from decisive refutation evidence.
This check establishes whether the passage occurs in the cached text.
Its relevance to the claimed overlap remains a matter of interpretation, and an unmatched passage leaves the textual support unverified.

The report combines target metadata, the core-task landscape, taxonomy, sibling comparisons, contribution-level analyses, and source references.
The scoring payload contains the structured comparison records and their narrative summaries.
The single-LLM prompt gives the structured records precedence when the two conflict.
Both scoring designs treat missing refutation, sparse retrieval, and placement in a singleton taxonomy leaf as insufficient evidence for a positive novelty claim.

\subsection{Multi-Agent Scoring}
\label{sec:scoring}
The scoring stage uses the five-level RINoBench rubric~\cite{schopf2026ideanovelautomatedbenchmark}.
Scores 1 and 2 represent existing work and minor variation; score 3 represents meaningful recombination, adaptation, or incremental extension; scores 4 and 5 represent new aspects and highly innovative directions, respectively.
Appendices~\ref{app:rubric} and~\ref{app:prompts} provide the full scale and selected scoring instructions.
The scorer selects the label best supported by the comparison records.

The workflow assigns Prosecutor, Defender, Judge, and Critic roles, following the use of explicit roles and communication protocols in multi-agent reasoning~\cite{yan2025selftalkcommunicationcentricsurveyllmbased}.
Table~\ref{tab:roles} summarises their inputs and outputs.
The Prosecutor and Defender are instructed to use only the structured scoring payload.
The Prosecutor identifies central overlap and minor variations; the Defender identifies supported technical distinctions or meaningful recombinations.
Each returns at most four short claims, a score-related signal, a rationale, and a qualitative confidence label.
The final score is assigned by the Judge.

\begin{table}[htbp]
  \caption{Roles in the fixed scoring workflow. All roles receive the structured evidence payload. Confidence labels are model outputs, not calibrated probabilities.}
  \label{tab:roles}
  \centering
  \small
  \renewcommand{\arraystretch}{1.15}
  \begin{tabularx}{\textwidth}{@{}p{0.14\textwidth}YY@{}}
    \toprule
    Role & Additional input & Responsibility and output \\
    \midrule
    Prosecutor & None & Evidence-supported overlap; low-novelty claims and signal. \\
    Defender & None & Specific central distinctions; high-novelty claims and signal. \\
    Judge & Both arguments; calibration exemplars & Score 1--5, justification, evidence, and adjacent-label explanations. \\
    Critic & Both arguments; Judge output; exemplars & Rubric audit; accept or revise, suggested score, and boundary-error reason. \\
    \bottomrule
  \end{tabularx}
\end{table}

The Judge receives the evidence payload, both arguments, and fixed exemplar cards illustrating the scale.
It outputs a score, confidence, a short justification, selected evidence points, and explanations for rejecting the adjacent lower and higher labels.
The cards serve as calibration examples, separate from the evidence about the target.
The prompt discourages using 3 as a default and permits extreme labels when supported.

The Critic checks the Judge's use of the scale for unsupported extremes, central-tendency bias, and errors at adjacent score boundaries.
It returns either \texttt{accept} or \texttt{revise}, together with a suggested score and the reason for any correction.
If revision is requested, the Judge reconsiders exactly once using the same payload, Prosecutor and Defender outputs, exemplar cards, and Critic feedback.
No new arguments are generated and no additional literature is retrieved during revision.
The process ends after this step.

\section{Evaluation}
\label{sec:evaluation}

\subsection{Data and Reference Labels}
\label{sec:data}
The scoring evaluation uses RINoBench, a benchmark for judging the novelty of research ideas with reference to related work, expert-derived scores, and textual justifications~\cite{schopf2026ideanovelautomatedbenchmark}.
Ten samples were randomly selected from each of the five score classes, giving a balanced set of 50 samples.
For each entry, NoveltyScope processes the corresponding paper PDF or URL and retains the benchmark's novelty label.

The labels originate in expert reviews of ICLR 2022 and ICLR 2023 submissions, for which reviewers had access to the papers.
Applying them here assumes that an idea-level label also describes the contributions extracted from the corresponding paper.
That assumption has not been separately validated.
All aggregate results below refer to the balanced label distribution, which can differ from the distribution in a set of naturally occurring submissions.

\subsection{Configurations and Metric}
\label{sec:configurations}
Table~\ref{tab:results} lists the five scoring configurations and their results.
E1 is a dense-vector retrieval baseline without an LLM: it retrieves similar papers and maps similarity to a novelty score.
It does not use the report-generation or multi-agent pipeline.

E2 and E3 use Gemini-2.5-flash, while E4 and E5 use Qwen3-Next-80B-A3B-Instruct, abbreviated as Qwen3 in the table.
The single-LLM conditions assign a score directly from NoveltyScope's structured evidence.
The multi-agent conditions use the Prosecutor, Defender, Judge, and Critic workflow.
The experimental setup specifies shared scoring evidence within each backend; per-sample payloads are not included in the accompanying materials.
Appendix~\ref{app:configuration} lists the generation and retrieval settings.

The metric is mean absolute error (MAE):
\begin{equation}
  \operatorname{MAE}=\frac{1}{N}\sum_{i=1}^{N}|\hat y_i-y_i|,
  \qquad N=50,
  \label{eq:mae}
\end{equation}
where $\hat y_i$ is the predicted score and $y_i$ the reference label.
MAE measures the average distance from the reference label in score units, treating adjacent labels as equally spaced.
This metric evaluates score agreement; report quality is considered separately in the case study.

\begin{table}[htbp]
  \caption{Scoring results on 50 class-balanced samples. The last column gives relative MAE reduction, $100(1.68-\mathrm{MAE})/1.68$, with E1 as the reference. All LLM conditions use NoveltyScope evidence.}
  \label{tab:results}
  \centering
  \small
  \renewcommand{\arraystretch}{1.15}
  \begin{tabularx}{\textwidth}{@{}lYllr@{}}
    \toprule
    Group & Scoring configuration & Backend & MAE $\downarrow$ & Reduction vs.\ E1 \\
    \midrule
    E1 & Dense similarity mapping & None & 1.68 & -- \\
    E2 & Single-LLM scoring & Gemini & 1.48 & 11.9\% \\
    E3 & Multi-agent scoring & Gemini & 1.26 & 25.0\% \\
    E4 & Single-LLM scoring & Qwen3 & 1.22 & 27.4\% \\
    E5 & Multi-agent scoring & Qwen3 & \textbf{1.16} & 31.0\% \\
    \bottomrule
  \end{tabularx}
\end{table}

\subsection{Quantitative Results and Interpretation}
\label{sec:results}
All four LLM configurations have lower MAE than the vector baseline.
Within the Gemini pair, MAE decreases from 1.48 to 1.26, an absolute difference of 0.22 and a relative reduction of 14.9\% from E2.
Within the Qwen3 pair, it decreases from 1.22 to 1.16, an absolute difference of 0.06 and a relative reduction of 4.9\% from E4.
E5 has the lowest MAE, 1.16, which is 0.52 below E1.
The Qwen3 conditions have lower error than the corresponding Gemini conditions on this sample.

The scoring prompts differ as well as the number of roles.
The stored single-LLM prompt prefers 3 over an extreme score when evidence is mixed, whereas the multi-agent Judge prompt discourages a middle-score default.
Their score-5 boundaries also differ: the single-LLM prompt requires both a distinctive core framing and mechanism, while the multi-agent prompt permits a highly distinctive framing or mechanism.
These differences can affect the errors and confound attribution to agent count.
Appendix~\ref{app:prompts} reproduces the relevant instructions from both stored templates.

Confidence intervals, significance tests, per-class errors, and repeated-run distributions are not reported; the repetition count is unspecified.
The results support a preliminary comparison of configurations on this sample.
The effects of retrieval fusion, adversarial roles, exemplars, and the Critic are not isolated by separate ablations.

\subsection{Illustrative Case Study: HiCLIP}
\label{sec:case}
The case study uses a report generated for HiCLIP: Contrastive Language-Image Pretraining with Hierarchy-aware Attention~\cite{geng2023hiclipcontrastivelanguageimagepretraining}.
The report identifies the core task as improving cross-modal alignment by incorporating hierarchy-aware attention into CLIP.
It lists two extracted claims, hierarchy-aware CLIP and a Group Transformer for vision, and records ten candidate comparisons for each.
Table~\ref{tab:case} summarises the two contribution-level analyses.

\begin{table}[htbp]
  \caption{System-generated comparisons for the two extracted HiCLIP contributions. The last column states the limits of each comparison.}
  \label{tab:case}
  \centering
  \small
  \renewcommand{\arraystretch}{1.15}
  \begin{tabularx}{\textwidth}{@{}p{0.23\textwidth}YY@{}}
    \toprule
    Extracted claim & Generated comparison & Scope of the comparison \\
    \midrule
    Hierarchy-aware CLIP & Tree Transformer is flagged for overlap in language-side hierarchical attention; paired passages are shown. & Component overlap does not alone refute the full multimodal contribution. \\
    Group Transformer for vision & No refutation is recorded among ten candidates; brief technical distinctions are provided. & The absence of refutation is limited to these ten candidates. \\
    \bottomrule
  \end{tabularx}
\end{table}

The report pairs a passage describing HiCLIP's relation to Tree Transformer with a candidate passage describing constraints that encourage tree-structured attention.
That comparison concerns the language-side mechanism.
Assessing the full contribution would also require considering the visual extension and multimodal integration.
For the vision contribution, the report instead presents short notes distinguishing retrieved candidates from the claimed grouping mechanism.

The report also contains a taxonomy, sibling comparisons, and source links.
Without a systematic human evaluation of passage faithfulness, comparison correctness, or usefulness, this case illustrates the output format but provides no estimate of general report quality or retrieval effectiveness.

\section{Discussion and Limitations}
\label{sec:discussion}

\paragraph{Evidence coverage and claim selection.}
The corpus determines which prior work the system can retrieve.
The inspected corpus snapshot contains 1167 records, including four dated 2024. The retrieval implementation does not filter candidates by the target paper's submission date, so the comparisons are not time-restricted.
Title-and-abstract indexing can miss mechanisms described only in technical sections, while semantic similarity can retrieve papers that share terminology but use different methods.
Extracting at most two claims from selected sections further restricts the assessment.
The resulting score applies to those claims and retrieved documents; changing either can change the judgement.

\paragraph{Traceability and error propagation.}
The HiCLIP example shows why a passage match and a valid refutation must be distinguished.
A quoted mechanism may overlap with one component while leaving the rest of a contribution unaddressed.
Errors in extraction, taxonomy construction, task comparison, contribution analysis, or narrative generation can propagate into scoring.
The Critic uses the same evidence as the Judge, so its revision is limited to the scoring decision.
Measuring report quality would require separate judgements of the selected comparisons, their evidence, and the generated explanations.

\paragraph{Evaluation validity.}
The evaluation covers 50 balanced samples and two model backends.
Its main validity concerns are the transfer of idea-level labels to full-paper inputs and the different scoring instructions in the paired conditions.
Per-class errors are needed to assess discrimination at the ends of the scale, which aggregate MAE alone cannot establish.
Repeated runs are needed to quantify variability, including at low temperatures.
Isolating the effect of the scoring roles would require a comparison with matched instructions and scoring payloads.

\paragraph{Resources and reproducibility.}
Processing time is reported as more than ten minutes per sample.
Latency distributions, hardware-normalised timings, and per-configuration monetary costs are unavailable.
Multiple scoring roles add calls and repeat the processing of evidence; token limits also occasionally truncated outputs passed to later stages.
These omissions limit independent reconstruction of the experiment.

\paragraph{Use as decision support.}
A novelty score addresses one aspect of a paper; correctness, usefulness, and overall scholarly value require other assessments.
Reviewers need the report and source passages to check the selected claims, challenge an overlap judgement, or identify missing prior work.
The score is intended to accompany that review.
Further work could extend corpus and full-text coverage, improve task-specific calibration, and test larger models or multi-round interaction, with report quality and computational cost evaluated alongside score agreement.

\section{Conclusion}
\label{sec:conclusion}
NoveltyScope retrieves related papers, compares their tasks and contributions with a target paper, and uses four scoring roles to assign a five-level novelty label.
Its report records the comparisons and source passages, as illustrated by the HiCLIP case.
On 50 balanced RINoBench samples, multi-agent scoring has lower MAE than single-LLM scoring under both tested backends.
The lowest MAE is 1.16, compared with 1.68 for the vector baseline.
Although computational and time constraints limit the evaluation of the multi-agent design's benefits, the study demonstrates the applicability to evidence-grounded novelty assessment.
The next steps are to document and extend corpus coverage, control the scoring instructions, and assess report quality separately from score agreement.

\bibliographystyle{IEEEtran}
\bibliography{references}

\clearpage
\appendix
\section{Configuration}
\label{app:configuration}
The embedding model is \texttt{all-MiniLM-L6-v2}, with L2-normalised vectors and dense-retrieval top-$k=30$.
RRF uses equal channel weights of 1 and fusion constant 60.
The evaluated language-model backends are Gemini-2.5-flash and Qwen3-Next-80B-A3B-Instruct.
Table~\ref{tab:configuration} summarises the available generation settings. In the current multi-agent implementation, the 2,000-token setting is a maximum-output configuration for each model call.

\begin{table}[htbp]
  \caption{Generation settings recorded for each stage. The scope of the multi-agent token limit is unspecified.}
  \label{tab:configuration}
  \centering
  \small
  \renewcommand{\arraystretch}{1.15}
  \begin{tabularx}{\textwidth}{@{}Yrr@{}}
    \toprule
    Stage & Temperature & Maximum tokens \\
    \midrule
    Contribution extraction & 0.0 & 4,000 \\
    Prior-work query generation & 0.0 & 3,000 \\
    Query-variant generation & 0.7 & 1,000 \\
    Core-task extraction & 0.0 & 2,000 \\
    Taxonomy construction & 0.0 & 20,000 \\
    Narrative generation & 0.0 & 5,000 \\
    Core-task comparison & 0.0 & 1,200 \\
    Contribution analysis & 0.0 & 8,000 \\
    Overall novelty assignment & 0.0 & 1,800 \\
    Single-LLM scoring & 0.0 & 3,000 \\
    Multi-agent scoring & 0.0 & 2,000 \\
    \bottomrule
  \end{tabularx}
\end{table}

\section{Five-Level Novelty Rubric}
\label{app:rubric}
Table~\ref{tab:rubric} gives the RINoBench-aligned scale used for scoring~\cite{schopf2026ideanovelautomatedbenchmark}.

\begin{table}[htbp]
  \caption{Five-level novelty rubric.}
  \label{tab:rubric}
  \centering
  \small
  \renewcommand{\arraystretch}{1.2}
  \begin{tabularx}{\textwidth}{@{}cY@{}}
    \toprule
    Score & Definition \\
    \midrule
    1 & The idea is not novel. All aspects already exist in prior work. \\
    2 & The idea is marginally novel. It represents only a minor variation of existing work. \\
    3 & The idea is somewhat novel. Aspects already exist in prior work. However, it might combine known approaches in new ways, apply them to new contexts, or propose incremental updates. \\
    4 & The idea is novel. It introduces new aspects not present in existing work. \\
    5 & The idea is highly innovative and novel. It is not present in existing work and potentially encourages new thinking or opens up new research directions. \\
    \bottomrule
  \end{tabularx}
\end{table}

\clearpage
\section{Selected Scoring Instructions}
\label{app:prompts}
The following excerpts reproduce the stored scoring templates with display indentation removed and lines wrapped.
Complete copies are included in \texttt{supplementary/}.
Their correspondence to the runtime versions has not been verified.

\subsection{Prosecutor and Defender}
The Prosecutor distinguishes supported overlap from missing negative evidence:
\par\noindent\begin{minipage}{\linewidth}
\begin{lstlisting}[style=prompt,firstline=47,lastline=51,gobble=4]
    - Treat "can_refute with evidence" as the strongest low-novelty signal.
    - Treat "can_refute without evidence" as a weaker but still relevant overlap signal.
    - Treat "cannot_refute" and "unclear" as absence of negative evidence, not as positive evidence of novelty.
    - Sparse retrieval coverage does not itself support novelty.
    - A singleton taxonomy group or weak structural comparison does not itself support novelty.
\end{lstlisting}
\end{minipage}\par
The Defender requires a specific positive distinction:
\par\noindent\begin{minipage}{\linewidth}
\begin{lstlisting}[style=prompt,firstline=107,lastline=111,gobble=4]
    - Do NOT claim novelty merely because refutation is missing.
    - Do NOT use sparse coverage or niche taxonomy placement alone as evidence of novelty.
    - Only output a high_novelty_claim when it points to a specific supported distinction.
    - If a claim supports score 5, explain what makes the work highly distinctive rather than merely new.
    - If no clearly differentiated technical property is supported, say so in the rationale.
\end{lstlisting}
\end{minipage}\par
Both roles return argument lists, a score-related signal, rationale, and confidence; the Judge assigns the score.

\subsection{Judge and Critic}
The Judge is instructed to use the full score range:
\par\noindent\begin{minipage}{\linewidth}
\begin{lstlisting}[style=prompt,firstline=156,lastline=162,gobble=4]
    - This is closest-label classification under the available retrieved evidence.
    - Your task is not to prove absolute novelty or absolute non-novelty.
    - Choose the score whose definition is best supported by the balance of evidence.
    - Use the full 1-5 scale.
    - Scores 1 and 5 are valid and should be used when supported.
    - Do not compress uncertain cases into 2-4.
    - Do not use score 3 as a safe default.
\end{lstlisting}
\end{minipage}\par
The score-5 criterion is:
\par\noindent\begin{minipage}{\linewidth}
\begin{lstlisting}[style=prompt,firstline=191,lastline=191,gobble=4]
    - Use score 5 when the central framing or mechanism is highly distinctive and not substantially covered by retrieved prior work.
\end{lstlisting}
\end{minipage}\par
The Critic either accepts the score or identifies a boundary-specific correction:
\par\noindent\begin{minipage}{\linewidth}
\begin{lstlisting}[style=prompt,firstline=272,lastline=274,gobble=4]
    - If decision is "accept", keep suggested_score equal to the Judge score.
    - If decision is "revise", suggested_score must be the closest corrected score.
    - The reason must identify the specific boundary error, such as "3 vs 4 boundary" or "2 vs 3 boundary".
\end{lstlisting}
\end{minipage}\par
\par\noindent\begin{minipage}{\linewidth}
A revision uses the same arguments and evidence:
\begin{lstlisting}[style=prompt,firstline=294,lastline=303,gobble=4]
    Rejudge exactly once using:
    - the same structured scoring payload
    - the Prosecutor output
    - the Defender output
    - the Critic feedback
    - the same fixed exemplar cards

    This is the only allowed revision.
    Do not rerun or invent new Prosecutor/Defender arguments.
    Finalize one score from 1 to 5.
\end{lstlisting}
\end{minipage}\par

\subsection{Single-LLM Comparison}
The single-LLM template uses structured evidence, but its instructions favour a middle score under mixed evidence and set a different score-5 criterion:
\par\noindent\begin{minipage}{\linewidth}
\begin{lstlisting}[style=prompt,firstline=71,lastline=73]
- If evidence is mixed and no exact-core overlap clearly dominates, prefer 3 over an extreme score.
- Do not assign 1 unless exact-core overlap covers most of the central mechanism.
- Do not assign 5 unless both the core framing and the core mechanism are positively supported as conceptually distinct.
\end{lstlisting}
\end{minipage}\par
The differences in these instructions are part of the configuration comparison discussed in Section~\ref{sec:results}.

\end{document}